\documentclass[usenatbib,useAMS,usegraphicx]{mn2e}
\usepackage{amssymb}
\usepackage{times}

\begin{document}

\title[Eccentric chaotic zone]
  {Dependence of a planet's chaotic zone on particle
eccentricity: the shape of debris disc inner edges}

\author[A. J. Mustill \& M. C. Wyatt]
  {Alexander J. Mustill$^{1,2}$\thanks{Email: alex.mustill@uam.es}, Mark C. Wyatt$^2$\\
  $^1$ Departamento de F\'isica Te\'orica, Universidad Aut\'onoma de Madrid, Cantoblanco, 28049 Madrid, Espa\~na\\
  $^2$ Institute of Astronomy, University of Cambridge, Madingley Road,
  Cambridge CB3 0HA, UK}

\maketitle

\begin{abstract}

The orbit of a planet is surrounded by a chaotic zone wherein nearby particles' orbits are chaotic and unstable. Wisdom (1980) showed that the chaos is driven by the overlap of mean motion resonances which occurs within a distance $(\delta a/a)_\mathrm{chaos}\approx 1.3 \mu^{2/7}$ of the planet's orbit. However, the width of mean motion resonances grows with the particles' eccentricity, which will increase the width of the chaotic zone at higher eccentricities. Here we investigate the width of the chaotic zone using the iterated encounter map and N-body integrations. We find that the classical prescription of Wisdom works well for particles on low-eccentricity orbits. However, above a critical eccentricity, dependent upon the mass of the planet, the width of the chaotic zone increases with eccentricity. An extension of Wisdom's analytical arguments then shows that, above the critical eccentricity, the chaotic zone width is given by $(\delta a/a)_\mathrm{chaos}\approx 1.8 e^{1/5}\mu^{1/5}$, which agrees well with the encounter map results. The critical eccentricity is given by $e_\mathrm{crit}\approx 0.21 \mu^{3/7}$. This extended chaotic zone results in a larger cleared region when a planet sculpts the inner edge of a debris disc comprised of eccentric planetesimals. Hence, the planet mass estimated from the classical chaotic zone may be erroneous. We apply this result to the HR~8799 system, showing that the masses of HR~8799~b inferred from the truncation of the disc may vary by up to 50\% depending on the disc particles' eccentricities. With a disc edge at 90\,AU, the necessary mass of planet~b to cause the truncation would be 8--10 Jovian masses if the disc particles have low eccentricities ($\lesssim0.02$), but only 4--8 Jovian masses if the disc particles have higher eccentricities. Our result also has implications for the ability of a planet to feed material into an inner system, a process which may explain metal pollution in White Dwarf atmospheres.

\end{abstract}

\begin{keywords}
chaos --- celestial mechanics --- planets and satellites: dynamical evolution and stability --- circumstellar matter --- stars: individual: HR~8799
\end{keywords}

\section{Introduction}

Planets strongly affect the dynamics of bodies on nearby orbits. Bodies coming within around one Hill's radius are strongly perturbed on the synodic time-scale and can either be scattered, resulting eventually in collision with the star or ejection from the system, or collide with the planet. However, beyond this region there is a chaotic zone of unstable orbits. Here, orbits diffuse through phase space and may eventually find themselves intersecting the planet's orbit, which will again result in a strong scattering or collision event. This unstable chaotic zone is thought to be responsible for sculpting the inner edges of debris discs such as that around Fomalhaut \citep{QuillenFom06,Kalas+08,Chiang+09}, with the location and shape of the disc edge being determined by the mass and location of the planet. The zone may also provide a reservoir for feeding bodies from an outer disc into the inner system. This is likely to be important in the late stages of stellar evolution, when scattering of planetesimals by planets has been invoked to explain metal pollution observed in some White Dwarf atmospheres as well as hot discs around them (\citealt{Zuckerman+03}; \citealt*{FarihiJura&Zuckerman09,BMW11}).

Chaos in dynamical systems is often driven by overlapping resonances \citep{Chirikov79}. \cite{Wisdom80} derived an analytical formula for the extent of a planet's chaotic zone based on overlapping first-order mean motion resonances, finding
\begin{equation}\label{eq:chaoticzone}
(\delta a/a)_\mathrm{chaos}=c\mu^{2/7}
\end{equation}
where $\mu=m_\mathrm{pl}/m_\star$ is the ratio of planetary to stellar mass and the coefficient $c=1.3$. This has been confirmed numerically, both using iterations of the encounter map \citep[][see \S\,\ref{sec:chaos width} below]{DQT} and in full N-body integrations \citep{Chiang+09}, although the coefficient $c$ when determined numerically is somewhat higher. However, the result of \cite{Wisdom80} is only valid for low-eccentricity particles, since the widths of mean motion resonances increase at larger eccentricities \citep{MD99}. \cite{Quillen&Faber06} extended Wisdom's result to consider the behaviour of particles orbiting close to an eccentric planet with low free eccentricities, which is relevant for modelling the Fomalhaut disc. They showed that the width of the zone for particles orbiting close to an eccentric planet with low free eccentricities is the same as that for low-eccentricity particles orbiting close to a circular planet.

 \begin{figure*}
\includegraphics[width=0.48\textwidth]{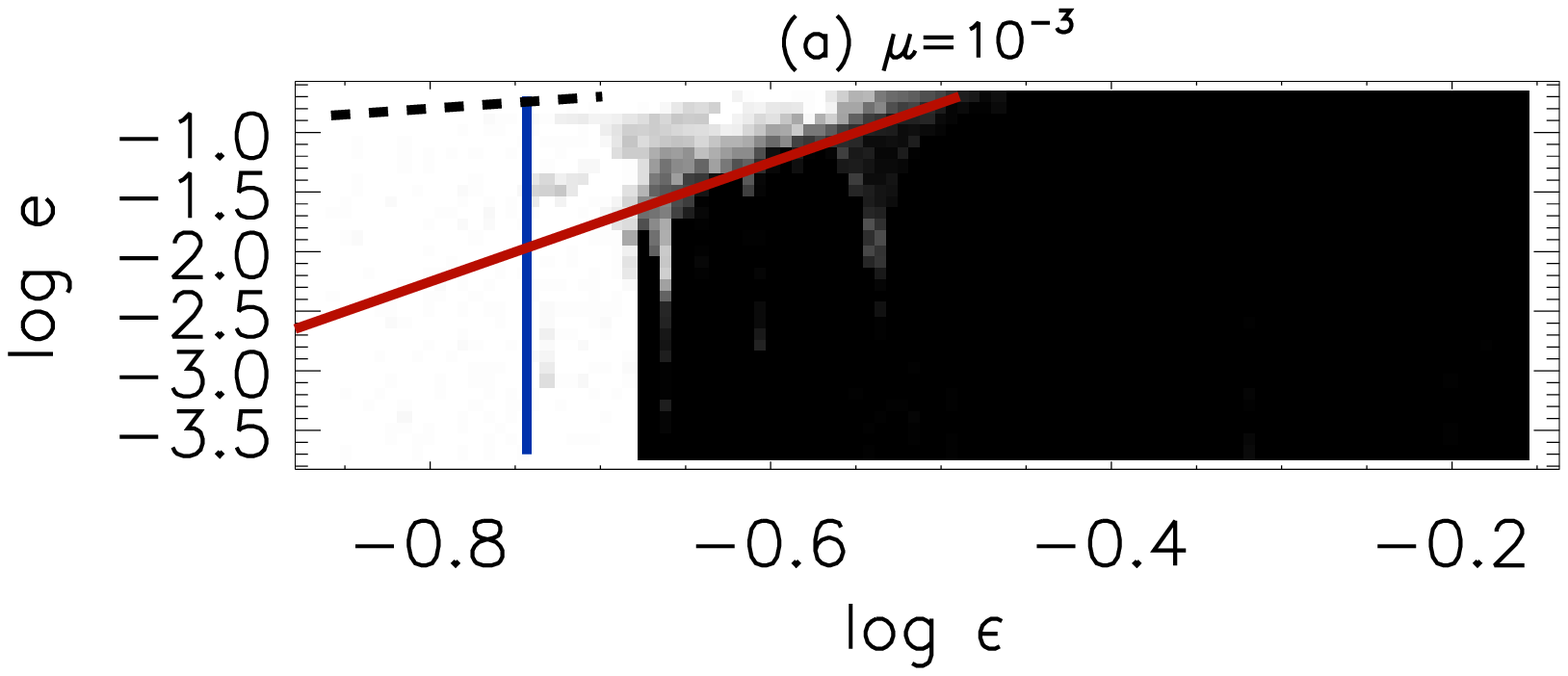}
\includegraphics[width=0.48\textwidth]{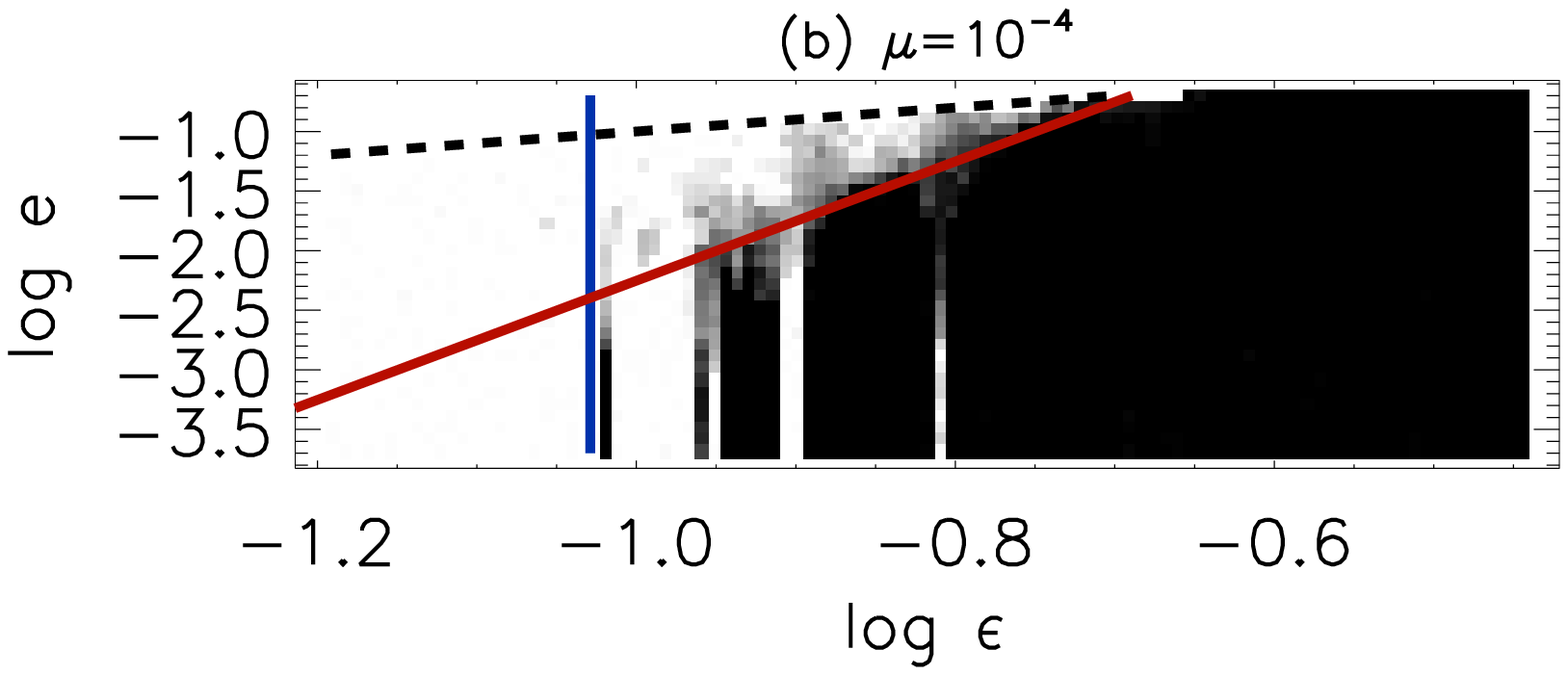}
\includegraphics[width=0.48\textwidth]{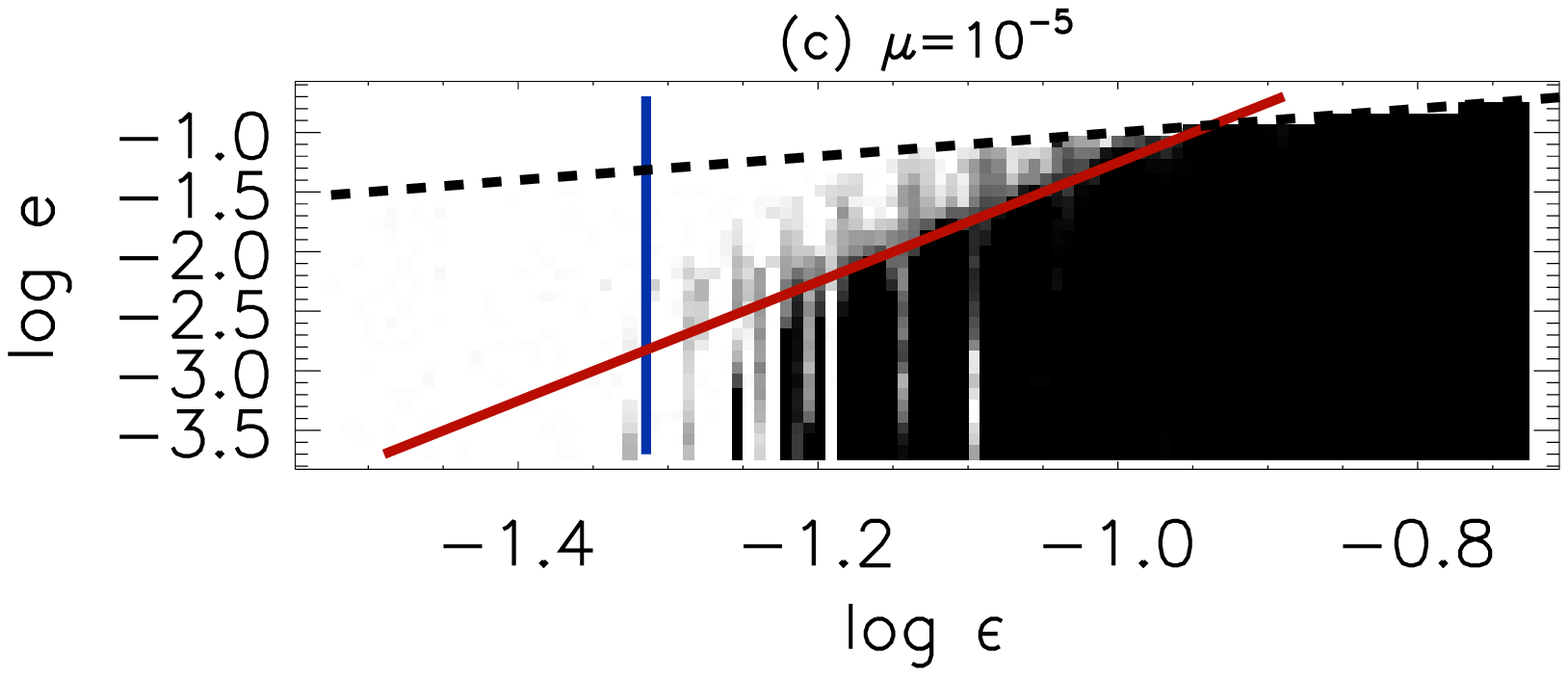}
\includegraphics[width=0.48\textwidth]{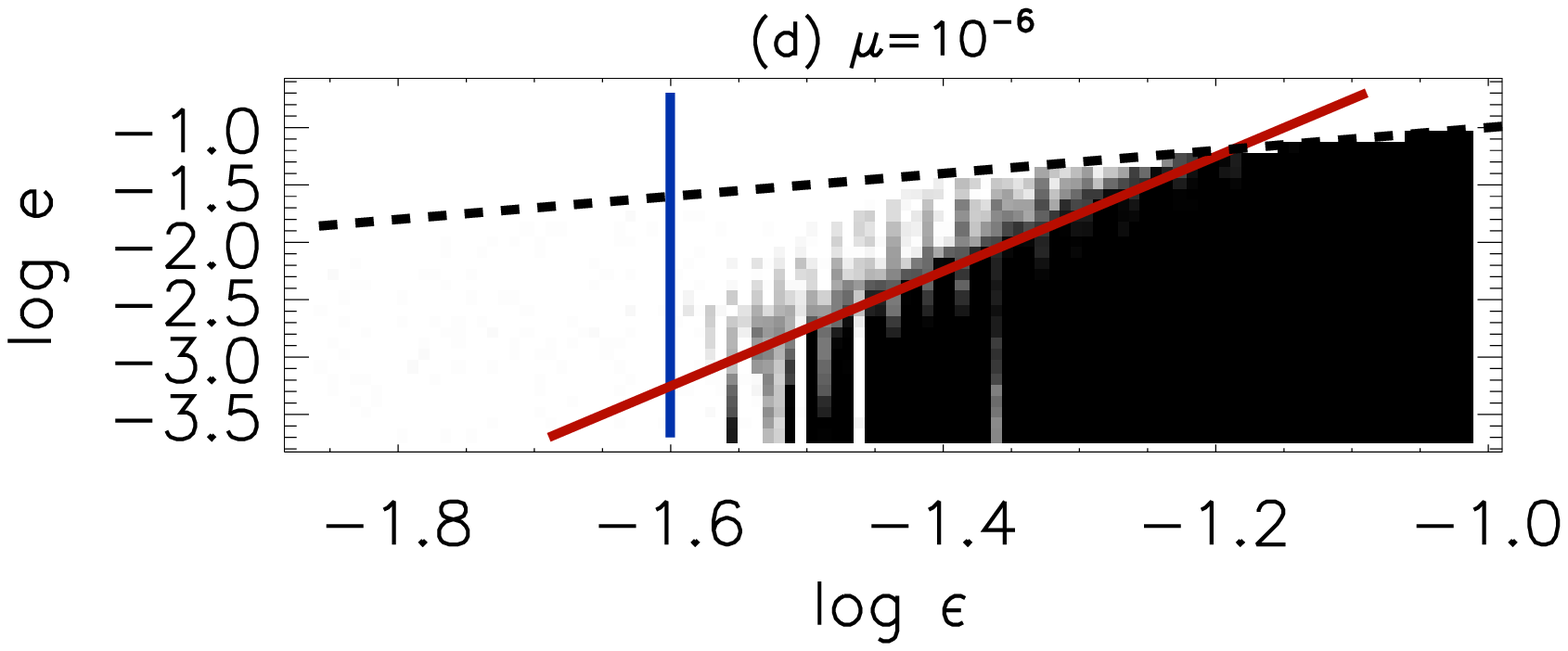}
\includegraphics[width=0.48\textwidth]{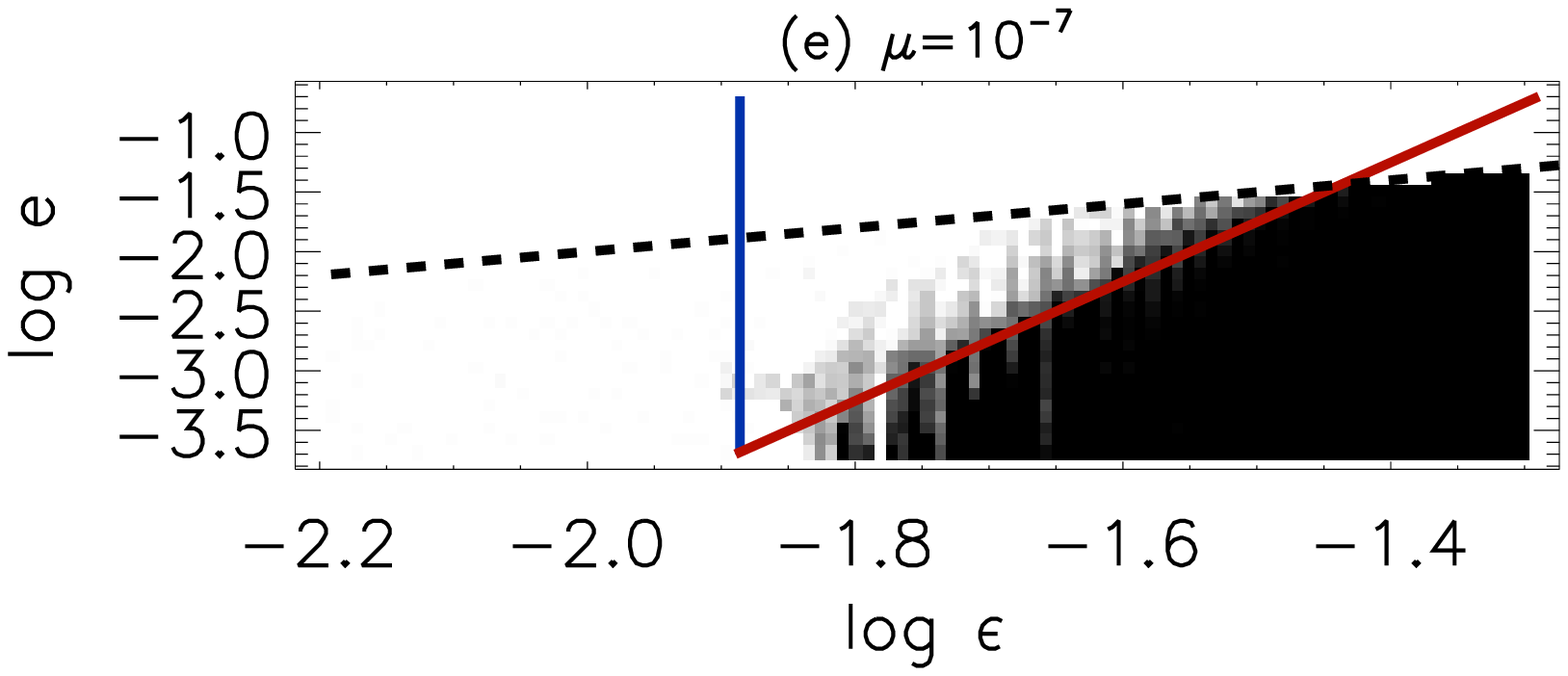}
\includegraphics[width=0.48\textwidth]{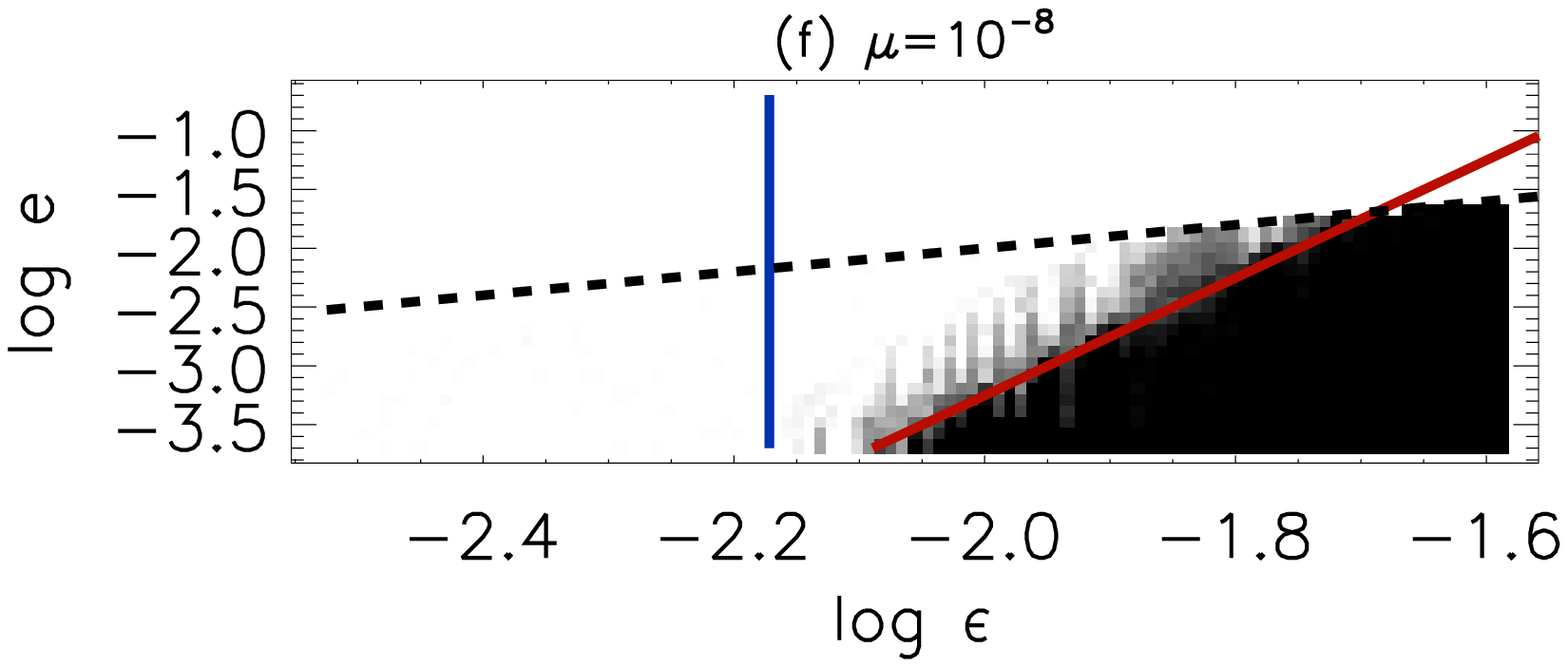}
\caption{\textbf{(a)--(f): }Extent of chaotic zone as a function of particle eccentricity and semi-major axis and planet mass. The right-hand side of each plot is at two Hill radii from the planet. The vertical blue line shows the extent of the classical chaotic zone (Wisdom, 1980). Above the dashed black line, the particles are on planet-crossing orbits; these particles were all classed as chaotic regardless of the map behaviour since they will experience close encounters with the planet. The solid red line shows the analytical estimate of the chaotic zone width as a function of eccentricity (Equation~\ref{eq:eccentric chaotic zone}).}
\label{fig:chaoticzone}
\end{figure*}

The effect the particles' free eccentricities have on the width of the chaotic zone has not been well explored. Since the Kuiper Belt Objects of our own Solar System can have free eccentricities of 0.1 and above, one might expect the debris discs of some extra-Solar systems to be stirred to a similar degree. \cite{BMW11} found, in N-body integrations, that the width of the chaotic zone for more eccentric particles ($e$ up to $0.1$) appeared to be greater than that for less eccentric particles. The small number of particles in the N-body integrations did not permit definite conclusions to be drawn, but they confirmed the increasing width using the computationally more efficient iterated encounter map \citep{DQT}. \cite{BMW11} speculated that the increasing width of the zone were due to the particles' pericentres entering the chaotic zone. This however seems unlikely since the chaos is caused by resonance overlap and that depends on the particles' semi-major axes and not directly on their pericentres. Here we demonstrate that the increasing chaotic zone width is instead due to the increasing width of mean motion resonances as particles' eccentricities are increased. \cite{Quillen07} and \cite{Chiang+09} investigated the effects of non-gravitational effects such as collisions and radiation forces on the inner edge of the disc, and here we investigate the effects of the particles' eccentricity.

One reason why it is important to characterise this chaotic zone properly is that currently much interest revolves around the effects that perturbing planets can have on debris disc morphology, particularly with a view to determining or refining the parameters of known or suspected planets. Accordingly, the $\mu^{2/7}$ law has been used to constrain the mass and semi-major axis of planets that may be sculpting the inner edges of debris discs \citep[\emph{e.g.,}][]{Moerchen+11}, to verify the consistency of planet and disc parameters determined through other means \citep[\emph{e.g.,}][]{Su+09}, and to aid setting up and interpreting N-body integrations \citep[\emph{e.g.,}][]{Chiang+09} studying the sculpting of debris disc edges. Hence, refinements to and deviations from this law are of prime importance for understanding the interaction of planets and debris discs. Given the large uncertainties which attend determinations of masses of planets detected by direct imaging \citep{Kalas+08}, dynamical constraints are very valuable and use of accurate dynamical models should be made. In \S\,\ref{sec:HR8799} of this paper, we discuss the HR~8799 system and how mass estimates of planet~b, thought to be sculpting the inner edge of the cold debris disc, are affected by deviations from the $\mu^{2/7}$ law.

This paper is organised as follows. In \S\,\ref{sec:chaos width} we use the encounter map and N-body integrations to explore the chaotic zone as a function of particle eccentricity. In \S\,\ref{sec:anal chaotic zone} we derive the width of the chaotic zone for eccentric particles using a resonance overlap criterion. In \S\,\ref{sec:chaos discussion} we discuss the implications of this for studies of debris discs sculpted by planets, particularly the HR~8799 system, and planetary systems orbiting White Dwarfs.

\section{Numerical investigation}
\label{sec:chaos width}

We investigated the structure of the chaotic zone using the iterated encounter map described in \cite{DQT}. This uses the approximate solution to Hill's Equations calculated by \cite{HP86}, which assumes a moderate separation between the planet and particle, to calculate the impulsive change to a particle's orbital elements due to a conjunction with the planet every synodic period. We determined whether orbits were regular or chaotic by examining the Fourier Transform of the eccentricity evolution. Regular orbits have power spectra with a few well-defined peaks, while the more chaotic an orbit the more peaks its power spectrum displays (this is a consequence of the phase space of a chaotic dynamical system being densely filled with periodic orbits). We chose a critical number of peaks, a peak being defined as where the slope of the power spectrum changes from positive to negative, to decide between regular and chaotic trajectories. For the investigations below, with $10^4$ iterations, we found that $1000$ peaks distinguished regular and chaotic orbits over the whole range of parameters considered. This classification method is similar to the spectral number chaos indicator \citep{SpectralNumber}.

In this way we then investigated the nature of trajectories as a function of the particles' semi-major axis $\epsilon=(a-a_\mathrm{pl})/a_\mathrm{pl}$ and eccentricity $e,$ and the planet:star mass ratio $\mu$. The results are shown in Figure~\ref{fig:chaoticzone}. In these plots, each point in the grid represents 100 particles, with initially random values of longitude of pericentre and mean anomaly. The greyscale shows the fraction of particles whose orbits were classed as chaotic, with black representing 100\% regular and white 100\% chaotic. In the region at the top left corner above the black line the particles are on planet-crossing orbits and hence the encounter map is invalid. These particles were classed as chaotic since they will be unstable anyway except perhaps for a few protected from close encounters by resonances.

From Figure~\ref{fig:chaoticzone} we see that the edge of the chaotic zone is a function of eccentricity, with the more eccentric particles being chaotic at greater separations from the planet than their less eccentric brethren. There are two regimes: a low-eccentricity regime in which the chaotic zone is independent of eccentricity, and a moderate-eccentricity regime where the width of the chaotic zone increases with eccentricity. The former regime is that identified by \cite{Wisdom80}. The critical eccentricity separating these regimes increases with $\mu$. The width at low eccentricity is in good agreement with the result from the resonance overlap criterion \citep[shown by a vertical blue line]{Wisdom80}, although this systematically underestimates it, as previous studies have shown \citep{DQT}. The increased chaotic zone width at larger eccentricities is likely due to the increasing width of the mean motion resonances at higher eccentricities, which can clearly be seen as intrusions of chaotic motion into the regions of regular motion. This increasing width can be derived analytically, as shown below.

\begin{figure}
  \includegraphics[width=.5\textwidth]{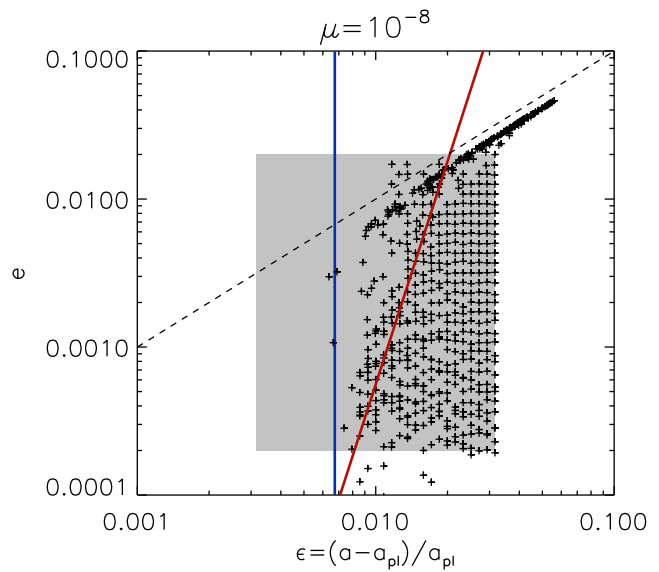}
  \caption{Fates of 961 particles after after a 10\,Myr N-body integration. Initially they are distributed over the grey rectangle and their final semi-major axes and eccentricities are plotted as crosses. Also shown are the same lines as in Figure~1: Dashed black: planet-crossing orbits; Solid blue: classical chaotic zone; Solid red: extended chaotic zone. Dynamical clearing above the red line is apparent.}
  \label{fig:nbody}
\end{figure}

To verify the accuracy of the encounter map results, we ran an N-body integration using the hybrid algorithm included in the \emph{Mercury} package \citep{ChambersMercury}. 961 test particles were placed on orbits exterior to a planet of mass $\mu=10^{-8}$ and their orbits integrated for 10\,Myr. These parameters correspond to Panel~(f) of Figure~\ref{fig:chaoticzone}. Their final eccentricities and semi-major axes are shown in Figure~\ref{fig:nbody}. Initially they fill the shaded region, but many are removed from regions closer to the planet and scattered onto orbits which almost intersect the planet's (shown by the dashed black line). The vertical solid blue line shows the chaotic zone width according to \cite{Wisdom80}. More eccentric particles are destabilised at greater semi-major axes than less eccentric ones, as was seen in the results from the encounter map. Hence, the encounter map appears to be giving an accurate picture of the dynamics. Additionally, the region above the dashed black line is almost totally devoid of particles, due to scattering by the planet (these particles have either collided with the planet or are in the unstable population just below the black line which is in the process of being scattered). This justifies our labelling of planet-crossing orbits as ``chaotic'' in Figure~\ref{fig:chaoticzone}.

\section{Analytical derivation}

\label{sec:anal chaotic zone}

\begin{figure}
\includegraphics[width=0.5\textwidth]{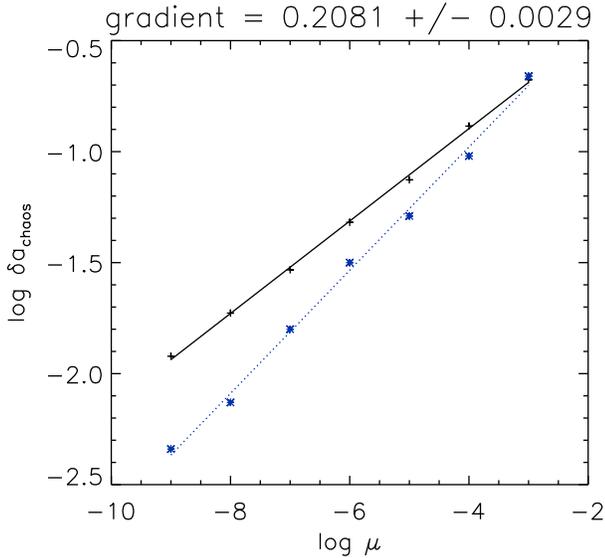}
\caption{\textbf{Black, solid line and crosses: }The chaotic zone width at a moderate eccentricity of $e=0.01$ as a function of planet:star mass ratio. Points show the numerical results from the grids in Figure~\ref{fig:chaoticzone} and the line shows the regression line, given by $(\delta a/a)_\mathrm{chaos}=0.86\mu^{0.208}$, close to the analytical result predicted by Equation~\ref{eq:eccentric chaotic zone}. \textbf{Blue, dotted line and stars: }The chaotic zone width at zero eccentricity. Here the regression line (dotted) has a slope of $0.278\pm0.0077$, close to the $2/7$ predicted by the analytical arguments of Wisdom (1980).}
\label{fig:width e=0.01}
\end{figure}

The standard derivation of the chaotic zone width \citep{Wisdom80,Quillen&Faber06} takes the widths of the mean motion resonances to be independent of eccentricity. Working from the resonance Hamiltonian
\begin{equation}\label{eq:ham}
\mathcal{H}=J^2+\beta J - J^{1/2}\cos\theta,
\end{equation}
where $J\propto e^2$ is the canonical momentum, $\theta$ is the resonant argument, and $\beta$ is a parameter measuring distance to the nominal resonance location \citep[see][]{QuillenRes06,MW11}, one can show that the width of a first-order resonance at low eccentricity is approximately
\begin{equation}
(\delta a/a)_\mathrm{res} = 3.2 \mu^{2/3}j^{1/3}
\end{equation}
\citep{Wisdom80}. Here $j$ numbers the resonance, $j:(j-1)$. In the limit of large $j$, adjacent resonances are separated by a distance
\begin{equation}
(\delta a/a)_\mathrm{sep}= 2/(3j^2).
\end{equation}
Equating these two distances gives the outermost resonance where overlap occurs as
\begin{equation}
j=0.51 \mu^{-2/7}.
\end{equation}
Since $a-a_\mathrm{pl}= 2/(3j)$ for large $j$, the chaotic zone width is thus given by
\begin{equation}\label{eq:circular chaotic zone}
(\delta a/a)_\mathrm{chaos}=1.3 \mu^{2/7}.
\end{equation}
This scaling is successfully reproduced by numerical iterations of the encounter map \citep[and Figure~\ref{fig:width e=0.01} below]{DQT}.

However, the resonance width grows with particle eccentricity, as can be seen in panel \textbf{(a)} of Figure~\ref{fig:chaoticzone}. This means that resonances that do not overlap at low eccentricities may overlap at higher eccentricities, and cause the chaotic zone to be wider. When eccentricities are not small, the width of the resonance can be found by considering the width of the resonant libration region after the resonant separatrix forms, which is given by $\delta J =4(4 J)^{1/4}$ \citep{MD99} in terms of the canonical momentum $J$. Now $J$ is related to $e$ by
\begin{equation}
J=k_j(m_\mathrm{pl}/m_\oplus)^{-2/3}(m_\star/m_\odot)^{2/3}e^2,
\end{equation}
where $k_j\sim(3jm_\odot/2m_\oplus)^{2/3}/2$ in the limit of large $j$ \citep{MW11,Quillen&Faber06}, and so
\begin{equation}
\delta J=1.3 \mu^{-2/3}j^{2/3}e\delta e.
\end{equation}
Using the relationship between eccentricity change and semi-major axis change in a resonance $\delta a = 2jae\delta e$ \citep{MD99}, we then have
\begin{equation}
(\delta a/a)_\mathrm{res} = 7.78 e^{1/2}\mu^{1/2}j^{1/2}
\end{equation}
for the maximum libration width in semi-major axis. Equating this to the resonance separation and redoing the previous derivation then gives
\begin{equation}\label{eq:eccentric chaotic zone}
(\delta a/a)_\mathrm{chaos}= 1.8  e^{1/5}\mu^{1/5}
\end{equation}
as the width of the chaotic zone at higher eccentricities. Note the two implications of this equation: first, the chaotic zone width grows weakly with eccentricity. The dependence is shown in Figure~\ref{fig:chaoticzone} as solid red lines. The agreement with the results from the encounter map is striking. We also show the dependence on Figure~\ref{fig:nbody}, where we can see many unstable particles above the solid red line. Second, the $\mu^{2/7}$ dependence of the chaotic zone width on planet mass, valid in the low-eccentricity regime, is replaced with a $\mu^{1/5}$ dependence. To verify this, in Figure~\ref{fig:width e=0.01} we plot the width of the chaotic zone at an eccentricity of $e=0.01$ as a function of $\mu$. The width was determined by finding the innermost cell where all trajectories were regular from the grids presented in Figure~\ref{fig:chaoticzone}. The regression line through these points gives a dependence $(\delta a/a)_\mathrm{chaos}=0.86\mu^{0.208}$, in excellent agreement with the analytical result from Equation~\ref{eq:eccentric chaotic zone}. Thus, for higher eccentricities, the chaotic zone width is given by a $\mu^{1/5}$ scaling, not the more familiar $\mu^{2/7}$. For comparison, the width of the chaotic zone at $e=0$ as a function of $\mu$ is shown on the same Figure. The regression line here gives $(\delta a/a)_\mathrm{chaos}=1.35\mu^{0.278}$, in excellent agreement with Wisdom's original result.

Equation~\ref{eq:eccentric chaotic zone} also allows the critical eccentricity separating the two regimes to be estimated. Equating $(\delta a / a)_\mathrm{chaos}$ obtained from Equation~\ref{eq:eccentric chaotic zone}, for eccentric particles, with the value from Equation~\ref{eq:circular chaotic zone}, for low-eccentricity particles, gives
\begin{equation}\label{eq:ecrit}
e_\mathrm{crit}\approx 0.21\mu^{3/7},
\end{equation}
growing with $\mu$, as is seen in Figure~\ref{fig:chaoticzone}. Note that $e=0.01$ is the critical eccentricity for $\mu=10^{-3},$ and indeed the widths of the classical and extended zones coincide at this point (Figure~\ref{fig:width e=0.01}).

Note that there is a maximum eccentricity for which this derivation of the extended chaotic zone is valid. This occurs where the solid red line in Figure~\ref{fig:chaoticzone} marking the boundary of the extended chaotic zone intersects the dashed black line denoting the region of planet-crossing orbits. At eccentricities above this, particles will be removed directly by close encounters rather than via chaotic diffusion. The maximum eccentricity is given by
\begin{equation}\label{eq:emax}
  e_\mathrm{max}=2.1\mu^{1/4},
\end{equation}
and in this regime particles are unstable if
\begin{equation}
  \delta a/a < e.
\end{equation}
In the next Section we adopt the slightly more conservative condition that particles are also unstable if they approach within one Hill's radius of the planet.

\section{Discussion}
\label{sec:chaos discussion}

\begin{figure}
  \includegraphics[width=.5\textwidth]{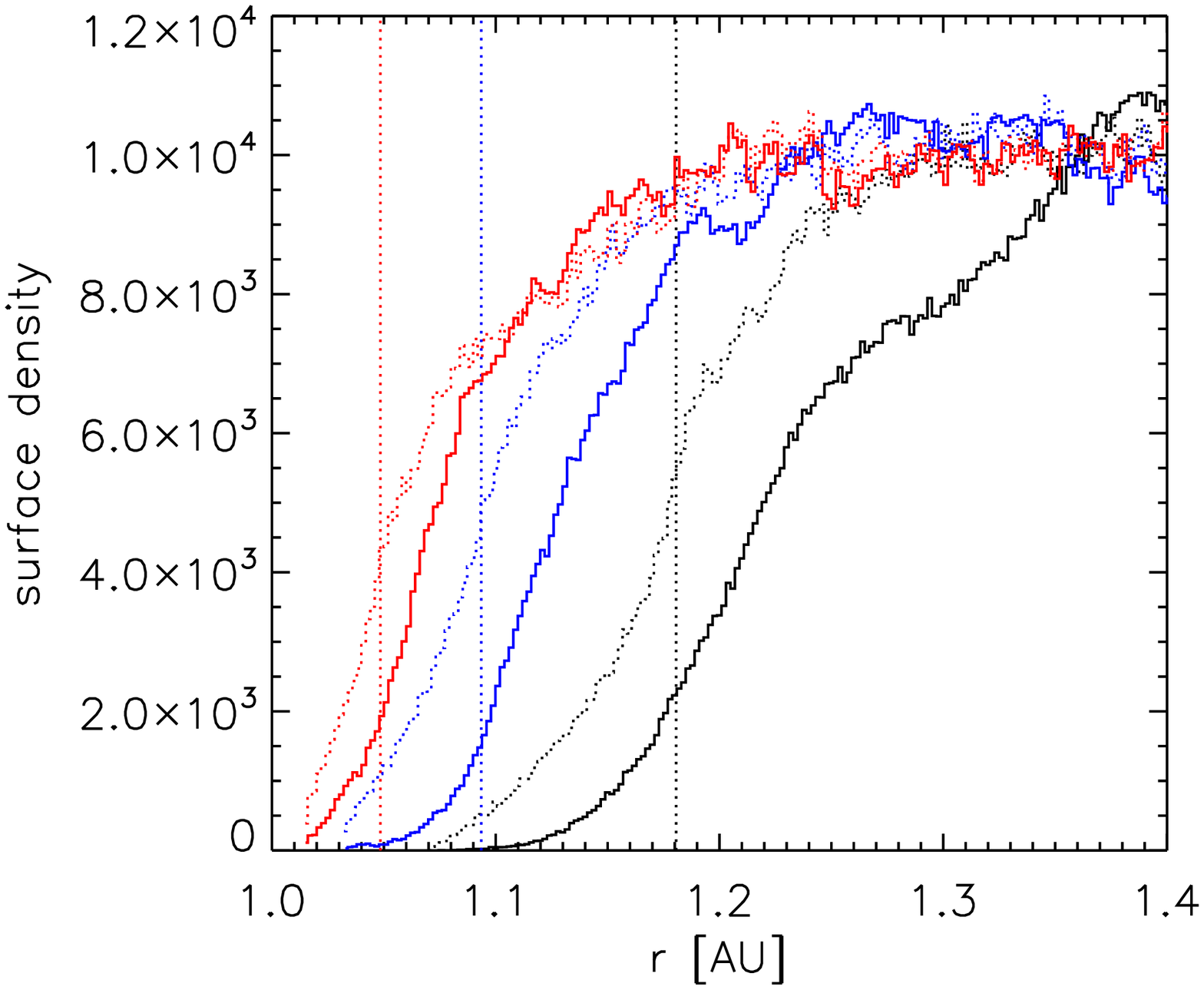}
  \includegraphics[width=.5\textwidth]{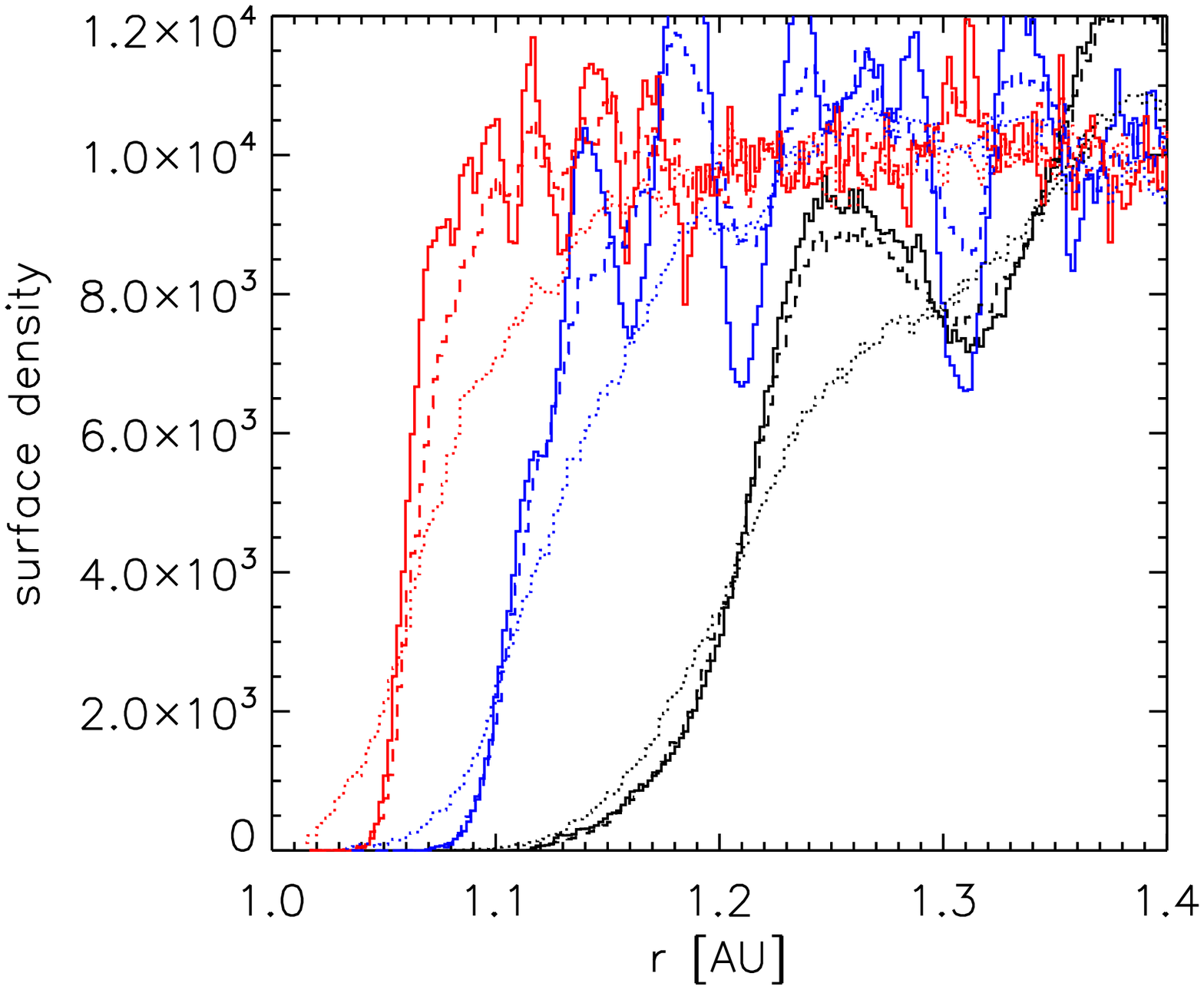}
  \caption{Surface densities at the inner edges of debris discs cleared by a planet at 1\,AU. \textbf{Top: }Comparison of the encounter map results (solid) with the prescription of Wisdom (1980; dotted). Vertical dotted lines show the location of the  classical zone edge in semi-major axis; irregular dotted lines show the planetesimal distribution when their eccentricities are accounted for. Maximum particle eccentricities are $0.1$. Planet masses are $\mu=10^{-3}$ (black), $\mu=10^{-4}$ (blue) and $\mu=10^{-5}$ (red). \textbf{Bottom: }Disc profiles according to the encounter map results for different planet masses and planetesimal eccentricities. Planet masses are as above; maximum particle eccentricities are 0.01 (solid), 0.03 (dashed) and 0.1 (dotted).}
  \label{fig:edge}
\end{figure}

\subsection{The shape of debris disc inner edges}

This result has important implications for the interactions of planets and debris discs. Discs showing inner clearings are commonly explained by the existence of a planet removing particles from its chaotic zone \citep[\emph{e.g.,}][]{QuillenFom06}. The mass and semi-major axis of the planet determine both the location and the sharpness of the inner edge \citep{Quillen&Faber06,QuillenFom06,Chiang+09}. Since our investigations above have revealed that the extent of the chaotic zone differs from the classical prescription, we now explore its effect on the nature of inner edges to debris discs. As discussed by \cite{Chiang+09}, to correctly model the shape of a disc edge it is necessary to account not only for particles located in semi-major axis at the edge of the chaotic zone, but also ones further out whose eccentricities may cause them to affect the radial distribution of particles at the disc edge. Accordingly, we generate surface density profiles for discs in the following way. We consider simple model discs where $5\times 10^4$ planetesimals are uniformly distributed in eccentricity between 0 and a maximum $e_\mathrm{m}$ and semi-major axis between 1\,AU and 2\,AU. Particles are removed if their orbits under iterations of the encounter map are chaotic, or if their periapsis comes within one Hill's radius of the planet. We then sample 100 mean anomalies from each particle's orbit, and sum the particles in radial bins to find the surface density profile. We also vary the particles' eccentricities and semi-major axes according to the encounter map results, because the particles do not remain at their initial $a$ and $e$ even if their orbits are regular.

The results, for mass ratios $\mu=10^{-3}$, $10^{-4}$ and $10^{-5}$ and a maximum eccentricity of $e_\mathrm{m}=0.1$, are shown in the top panel of Figure~\ref{fig:edge} as solid lines. We see that the shape of the inner disc edge depends on both the planet's mass, with more massive planets making the edge shallower. We show on the same plot with dotted lines the results from the original prescription of \cite{Wisdom80}, showing a discrepancy both in the location of the edge and in the shape of it. The discrepancy is greater for larger planet mass. In the bottom panel of Figure~\ref{fig:edge}, we show the results from the encounter map iterations for a range of planet masses and particle eccentricities. More massive planets and more eccentric particles lead to shallower slopes. In addition, fluctuations in the surface density outside the cleared zone caused by mean motion resonances can be seen \citep[see also][]{Quillen07}, and these are more apparent for lower eccentricities. The shallowness of the edge means that its precise location is ill defined, but if it is attempted to be quantified by a single statistic, for example the half-maximum, then it should be noted that this value depends on the particles' eccentricities, leading to a degeneracy between planet mass and particle eccentricity.

In this subsection, and the subsequent, we consider only the gravitational influence of the planet on the planetesimals which the disc comprises. It is likely that in many debris discs the eccentricities of planetesimals are high, if they have been dynamically excited by planets' secular perturbations \citep{MW09} or by large planetary embryos embedded in the disc \citep{KB10}. In such a disc collisions between planetesimals will be destructive due to the high relative velocities. In such a disc eccentricities will remain high as ongoing stirring processes will dominate over collisional damping \citep{KB04,SW11}. We do note that there may be discs where the relative velocities remain low \citep{Heng&Tremaine} in which eccentricities may be continually damped by collisions. The truncation of such discs will be governed by both dynamical clearing from the classical chaotic zone and viscous spreading of the disc due to the inelastic collisions \citep{Quillen07}. We do not consider such discs here.

\subsection{Case study: HR 8799}

\label{sec:HR8799}

As an example of the effects of the extended chaotic zone, consider the HR~8799 system. The 1.5 Solar mass star is orbited by 4 known planets and at least two debris belts. \cite{Su+09} estimated that the inner edge of the outer disc lies at 90\,AU based on fitting the disc's spectral energy distribution (SED); the inner edge of the disc is not resolved. It is thought that the outermost planet, HR~8799~b, at a projected distance of 68\,AU from the star \citep{Marois+08}, is responsible for the truncation. \cite{Su+09} argued that the radius of the truncation, the location of planet~b at 68\,AU, and a planet mass of 10 Jovian masses, at the upper end of the range suggested by photometry \citep{Marois+08}, are all consistent with a scenario in which the truncation is due to planet~b's chaotic zone. Quantitatively, \cite{Su+09} took the chaotic zone to be given by $1.4\mu^{2/7}$ as given by \cite{Malhotra98}.

\begin{figure}
  \includegraphics[width=0.5\textwidth]{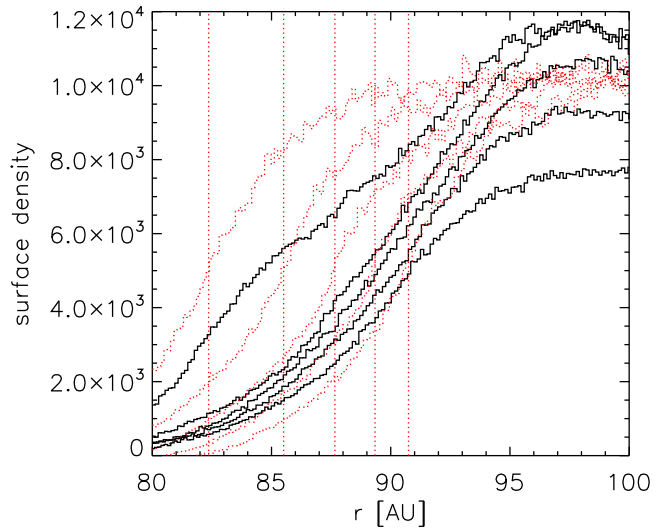}
  \caption{Comparison of disc edge profiles for the disc around HR~8799 truncated by planet~b at 68\,AU. The maximum eccentricity of the disc particles is $0.1$. Solid lines show disc profiles derived from the encounter map results; dotted show profiles when truncated by the classical Wisdom prescription. The vertical dotted lines show the location of the disc edge for particles on perfectly circular orbits. Planet masses increase from left to right: 2, 4, 6, 8 and 10 Jovian masses. Note that all the disc surface densities are $10^4$ far from the planet; the different heights visible here are due to the effects of the 2:1 resonance at 108\,AU.}
  \label{fig:HR8799}
\end{figure}

As we have seen, this relationship underestimates the width of the chaotic zone at high eccentricities. If the disc particles' eccentricities are above $\sim0.02$ (estimated from Equation~\ref{eq:ecrit}), the extent of the clearing will be greater than this because of the extended chaotic zone. Hence, if the disc particles have low eccentricities, the mass estimated by \cite{Su+09} will be correct, and the planet mass will be around $10$ Jovian masses. However, if the disc particles' eccentricities are higher than $\sim0.02$, the same clearing will be achieved with a smaller planet mass. We quantify this effect in Figure~\ref{fig:HR8799}. This compares the shape and extent of the clearing due to planets of 2, 4, 6, 8 and 10 Jovian masses, under the assumptions that particles are removed only in the classical chaotic zone (dotted lines) or also in the extended chaotic zone (solid line). In each case, particle eccentricities were distributed uniformly between $0$ and $0.1$. We also show as vertical dotted lines the location of the edge for particles on circular orbits. We see that the profiles for 8 and 10 Jovian mass planets under the Wisdom prescription (whose medians straddle 90\,AU) are similar to the profiles for the 4 and 8 Jovian mass planets, respectively, under the extended zone prescription. Hence, the planet masses required to achieve a given clearing may be less than predicted by the Wisdom criterion by as much as 50\%.

Thus, if the edge of the outer disc is located at 90\,AU, and the particles in the disc have eccentricities of order $0.1$, then the mass of planet~b would be in the range 4--8 Jovian masses, rather than the 8--10 Jovian masses that it would be if the particles' eccentricities were below $0.02$. For now it is not possible to make such firm conclusions, since observations have yet to resolve the inner edge of the disc, and its 90AU location inferred from SED modelling is degenerate with the assumptions made about the particle properties. Nevertheless, the inner edge will likely be resolved by future observations, and this will help to reduce the uncertainty in the planet mass. For example, some authors, \emph{e.g.,} \cite{Currie+11}, claim a lower mass for planet~b of 6--7 Jovian masses, and this is consistent with the planet truncating a disc of planetesimals, whose eccentricities range up to $0.1$, at 90\,AU, if the edge indeed be located there. The eccentricities of the disc particles may have been excited to such an extent by the growth of large planetesimals in the disc \citep{KB10}, secular perturbations from the planets, if one or more is eccentric \citep{MW09}, or sweeping by mean motion or secular resonances during past evolution of the system \citep{MW11,MM11}.

Finally, we note that our main conclusion -- that the mass of planet~b estimated from the extent of the chaotic zone will depend on the eccentricities of the disc particles -- does not depend on the actual location of the disc edge. If the disc prove to be at a greater radius, the mass estimated for a disc of eccentric particles will still be greater than that estimated for particles on circular orbits, although each will be greater than its corresponding mass estimated for a disc edge at 90\,AU.

\subsection{Post-Main Sequence evolution of planetary systems}

There are also important implications for the evolution of planetary systems when a star loses mass during post-Main Sequence evolution. When stars lose mass on the Asymptotic Giant Branch the planet:star mass ratios increase and previously stable systems can be destabilised \citep{Debes&Sigurdsson02,BMW11}. Stars typically lose mass at relatively modest rates, and under these conditions orbits expand adiabatically and the ratios of semi-major axes are unchanged \citep{Veras+11}. However, since the planet:star mass ratio increases as the star loses mass, quantities which depend on this ratio, such as the Hill's radius and the widths of resonances and the chaotic zone, will change too. The chaotic zone expands, and after mass loss particles which were previously on stable orbits may find themselves in the chaotic zone. The presence of this reservoir of newly unstable material may explain the existence of metal pollution in the atmospheres of White Dwarfs \citep{Zuckerman+03}, as well as hot discs orbiting some White Dwarfs \citep{FarihiJura&Zuckerman09}, as bodies are scattered onto highly eccentric orbits and tidally disrupted \citep{Zuckerman+03,BMW11}. \cite{BMW11} estimated the amount of destabilised material simply by expanding the $\mu^{2/7}$ zone according to the new, lower, stellar mass. While this will correctly describe the new chaotic zone at low eccentricity, seemingly vulnerable particles with higher eccentricity may escape being engulfed by the new chaotic zone. This is because the lower boundary of the extended chaotic zone increases as $\mu$ increases. However, a further complication may counteract this: \cite{Veras+11} showed that if mass loss rates are relatively rapid then particles' orbits do not expand adiabatically and they may acquire some eccentricity. This could push seemingly safe bodies into the extended chaotic zone. Since the eccentricities required to enter the extended chaotic zone are relatively modest, this could be an important mechanism for destabilising bodies. \\

\section*{Acknowledgements}

AJM is grateful for the support of an STFC studentship. We are indebted to Amy Bonsor for comments on the manuscript. We should like to thank the anonymous referee for suggesting improvements to the paper.

\bibliographystyle{mn2e}
\bibliography{bibliography}

\end{document}